\documentclass[pdflatex,sn-apa]{sn-jnl}

\usepackage{graphicx}%
\usepackage{multirow}%
\usepackage{amsmath,amssymb,amsfonts}%
\usepackage{amsthm}%
\usepackage{mathrsfs}%
\usepackage[title]{appendix}%
\usepackage{xcolor}%
\usepackage{textcomp}%
\usepackage{manyfoot}%
\usepackage{booktabs}%
\usepackage{algorithm}%
\usepackage{algorithmicx}%
\usepackage{algpseudocode}%
\usepackage{listings}%

\theoremstyle{thmstyleone}%
\theoremstyle{thmstyletwo}%

\theoremstyle{thmstylethree}%

\begin{document}

\title[Article Title]{SHAPR: A Solo Human-Centred and AI-Assisted Practice Framework for Research Software Development}


\author{\fnm{Ka Ching} \sur{Chan}}\email{kc.chan@unisq.edu.au}


\affil{\orgdiv{School of Business}, \orgname{University of Southern Queensland}, \city{Springfield Central}, \postcode{QLD 4300}, \state{Queensland}, \country{Australia}}





\abstract{Research software has become a central vehicle for inquiry and learning in many Higher Degree Research (HDR) contexts, where solo researchers increasingly develop software-based artefacts as part of their research methodology. At the same time, generative artificial intelligence is reshaping development practice, offering powerful forms of assistance while introducing new challenges for accountability, reflection, and methodological rigour. Although Action Design Research (ADR) provides a well-established foundation for studying and constructing socio-technical artefacts, it offers limited guidance on how its principles can be operationalised in the day-to-day practice of solo, AI-assisted research software development. This paper proposes the SHAPR framework—Solo, Human-centred, AI-assisted PRactice—as a practice-level operational framework that complements ADR by translating its high-level principles into actionable guidance for contemporary research contexts. SHAPR supports the enactment of ADR Building–Intervention–Evaluation cycles by making explicit the roles, artefacts, reflective practices, and lightweight governance mechanisms required to sustain human accountability and learning in AI-assisted development. The contribution of the paper is conceptual: SHAPR itself is treated as the primary design artefact and unit of analysis and is evaluated formatively through reflective analysis of its internal coherence, alignment with ADR principles, and applicability to solo research practice. By explicitly linking research software development, Human–AI collaboration, and reflective learning, this study contributes to broader discussions on how SHAPR can support both knowledge production and HDR researcher training.}




\maketitle

\section{Introduction}\label{intro}

Research software development has become an integral component of contemporary research practice, particularly in Higher Degree Research (HDR) contexts where scholars increasingly design and implement software artefacts as part of their methodological approach. At the same time, advances in generative artificial intelligence are reshaping how such software is developed, introducing new opportunities for acceleration alongside challenges related to accountability, learning, and research rigour. This section establishes the motivation for the study by examining research software as a knowledge-generating artefact, the emergence of AI-assisted solo development, and the suitability—and limitations—of existing research methodologies. It concludes by articulating the research aim and outlining the contributions of the paper.

\subsection{Research software as a knowledge-generating artefact}\label{intro1}

Research software has become central to scholarly inquiry across a wide range of disciplines, including information systems, education, data science, and cybersecurity \citep{Katz2021}. In many contemporary research contexts, software artefacts are no longer peripheral tools used solely for data processing or analysis; instead, they increasingly function as knowledge-generating artefacts \citep{Wilson2014} through which research questions are explored, theories are instantiated, and insights are produced. This shift is particularly evident in HDR projects, where doctoral, master’s, and honours students frequently design and implement bespoke software systems as part of their research methodology.

In such settings, software development is closely intertwined with learning, experimentation, and reflection. Building software artefacts enables researchers to externalise abstract ideas, test assumptions through implementation, and iteratively refine both the artefact and their conceptual understanding. From an educational perspective, this aligns with learning-by-building and constructionist views of knowledge development, where understanding emerges through active engagement with complex artefacts rather than passive consumption of information \citep{Papert1980,Kolb1984}. Within HDR contexts, research software thus serves a dual role: it is simultaneously a means of inquiry and an outcome of the research process \citep{Kennedy-Clark2013}.

A defining characteristic of research software development in HDR settings is that it is most often undertaken by a single researcher. Unlike industrial software projects, which typically involve teams with specialised roles and formal governance structures, research software is frequently developed by solo practitioners who must assume multiple roles, including domain expert, system designer, developer, evaluator, and reflective analyst. This solo mode of development affords flexibility and close integration between domain knowledge and technical design, but it also imposes significant cognitive and methodological demands. Decisions about system architecture, implementation trade-offs, and evaluation strategies are made by the same individual who is responsible for ensuring research rigour and scholarly contribution.

Despite its prevalence, solo research software development remains under-theorised in both software engineering and educational research literatures. Existing work on research software engineering often emphasises team-based practices, sustainability, and professional roles, while pedagogical studies frequently address project-based learning without explicitly considering software artefacts as instruments of research methodology. As a result, there is limited guidance on how solo researchers can structure software development activities in ways that systematically support learning, reflection, and knowledge generation.

This gap becomes increasingly consequential as research artefacts grow in scope and complexity. Many contemporary research systems are software-based and, in some cases, full-stack in nature, encompassing user interfaces, application logic, data persistence, and deployment environments. Developing such systems requires sustained design decisions across multiple layers, reinforcing the need for structured approaches that integrate technical work with reflective research practice \citep{Schon1983}.

At the same time, not all research artefacts take the form of full-stack systems. In many research contexts, artefacts may consist of algorithmic implementations, analytical or data-processing pipelines, simulation models, or formal analytical constructs such as statistical or structural equation models. In each case, inquiry is advanced through iterative construction, evaluation, and refinement rather than complete prior specification. Recognising research artefacts, software-based or otherwise, as knowledge-generating instruments therefore motivates the need for frameworks that explicitly support solo researchers in aligning development practice with rigorous research methodologies.

\subsection{Emergence of AI-assisted solo development}\label{intro2}

Recent advances in generative AI have begun to significantly reshape how research software is conceived and developed, particularly in solo research contexts. Large language models (LLMs) and AI-assisted development tools are increasingly capable of generating source code, drafting requirements, suggesting architectural patterns, and summarising technical artefacts \citep{Chow2025}. For solo researchers engaged in research software development, these tools offer the potential to accelerate implementation and reduce the technical barriers associated with breadth and depth of expertise \citep{Amershi2019,Shneiderman2020}.

One manifestation of this shift is the growing practice of so-called \emph{vibe coding}, in which developers rely on high-level prompts, iterative conversational feedback, and rapid experimentation rather than detailed upfront specifications \citep{Fawzy2025,Ge2025}. In research settings, such practices allow solo researchers to translate conceptual ideas directly into working prototypes within short timeframes. When combined with integrated development environments and version control platforms, AI-assisted tools can support rapid exploratory cycles that align well with the iterative nature of research inquiry. However, increased speed and abstraction also introduce challenges for maintaining methodological rigour and transparency.

AI-assisted development alters the cognitive distribution of work between human and machine. Tasks such as code generation, documentation drafting, and test suggestion may be partially delegated to AI systems, raising questions about authorship, accountability, and understanding. In research contexts, these questions are particularly salient because the researcher remains responsible not only for the correctness of the software artefact, but also for the validity and credibility of the knowledge claims derived from it. Uncritical reliance on AI-generated outputs risks obscuring design rationales, masking errors, and weakening traceability between research questions, design decisions, and outcomes.

From a learning perspective, the integration of AI into solo research software development presents both opportunities and risks. AI tools may function as cognitive scaffolds that support researchers in navigating unfamiliar technical domains, but they may also reduce opportunities for deep learning if outputs are accepted without reflection. This tension aligns with broader concerns in human--AI collaboration research regarding the balance between automation and meaningful human engagement \citep{Amershi2019}.

Despite rapid adoption in practice, existing research methodologies provide limited guidance on how AI-assisted tools should be systematically incorporated into research workflows. In particular, there is a lack of structured approaches that help solo researchers integrate AI assistance while preserving human accountability, reflective practice, and methodological coherence. As AI increasingly mediates research software development, frameworks are needed that explicitly align Human--AI collaboration with established research methodologies rather than leaving such integration to ad hoc practice.

\subsection{Why Action Design Research---and why it is insufficient on its own}\label{intro3}

Action Design Research (ADR) \citet{Sein2011} has emerged as a well-established methodology for studying and constructing socio-technical artefacts, particularly in information systems and educational technology research. By integrating artefact construction with intervention, evaluation, and reflection, ADR provides a structured approach for generating design knowledge while addressing real-world problems. Its emphasis on iterative Building--Intervention--Evaluation (BIE) cycles makes ADR well suited to research settings in which learning unfolds through the development and refinement of complex artefacts, including research software systems.

ADR offers several advantages in this context. It legitimises software artefacts as vehicles for inquiry rather than merely technical outputs, and it foregrounds reflection and abstraction as mechanisms for generating transferable design knowledge \citep{Mullarkey2019}. These characteristics align closely with the educational and knowledge-building goals of HDR research, where methodological rigour and reflective learning are central concerns.

However, while ADR provides a robust methodological foundation, it offers limited guidance on how its principles should be operationalised in day-to-day development practice, particularly in solo research contexts. The original formulation of ADR assumed organisational settings with multiple stakeholders, role separation, and collective sense-making processes---assumptions that do not readily translate to solo research software development.

This limitation is amplified when generative AI tools are introduced. ADR specifies what activities should occur, but it does not articulate how Human--AI collaboration should be structured within those activities, nor how accountability and understanding should be preserved when development tasks are partially delegated to AI systems. In the absence of explicit guidance, solo researchers may adopt practices that obscure decision-making, weaken traceability, and blur the boundary between human judgement and machine-generated output.

From an educational perspective, this gap has important implications. Without additional scaffolding, AI-assisted development risks prioritising rapid production over reflective learning and methodological transparency. While ADR establishes methodological intent, complementary guidance is needed to ensure that this intent is realised in practice when development is undertaken by a single researcher working with AI assistance \citep{Kennedy-Clark2013}.

These observations suggest that ADR, while necessary, is not sufficient on its own to address the practical and pedagogical challenges of AI-assisted solo research software development. Addressing this gap motivates the SHAPR framework proposed in this paper.

\subsection{Research aim and contributions}\label{intro4}

The aim of this study is to develop and articulate a human-centred framework that supports solo researchers in systematically enacting Action Design Research (ADR) when developing AI-assisted research software. Rather than proposing a new research methodology, the paper focuses on translating ADR’s methodological intent into actionable guidance that preserves human accountability, reflective learning, and research rigour in AI-mediated development contexts.

To achieve this aim, the paper introduces the SHAPR framework---a Solo, Human-centred, AI-assisted PRactice---positioned as a practice-level framework that complements ADR. SHAPR makes explicit how development activities, Human--AI roles, artefact management, and reflective processes can be structured within iterative Building--Intervention--Evaluation cycles. By foregrounding conceptual role separation and artefact-centred governance, the framework addresses the distinctive challenges faced by solo researchers who must assume multiple roles while working alongside generative AI systems.

This paper makes three contributions. First, it conceptualises solo research software development as a distinct and under-theorised research context, highlighting the methodological and educational implications of role multiplexing and AI-assisted practice. Second, it proposes SHAPR as transferable design knowledge \citep{Mullarkey2019} that operationalises ADR for AI-assisted research software development without constraining researchers to specific tools or technologies. Third, it clarifies the analytical relationship between research software artefacts and methodological scaffolding \citep{Alrawili2020} by treating SHAPR as the unit of analysis for this study.

The scope of the paper is intentionally conceptual. The SHAPR framework constitutes the primary artefact and unit of analysis, and it is evaluated through formative reflection on internal coherence, applicability to solo research contexts, and alignment with established ADR principles. No empirical instantiation of a specific research software system is reported. Instead, the paper establishes a foundation for applying SHAPR across multiple research software development projects in diverse disciplinary contexts, enabling cumulative investigation of AI-assisted research practice and learning.

\section{Background and Related Work}\label{background}

This section situates the study within the literature relevant to the development of the SHAPR framework. Rather than providing an exhaustive review, it focuses on strands of work that inform the methodological, technical, and educational context of the study. Specifically, the section reviews Action Design Research as a methodological foundation, examines research software and full-stack systems as sites of inquiry, considers the distinctive characteristics of solo research practice, and discusses Human–AI collaboration in software development. Together, these perspectives motivate the need for clearer practice-level guidance for contemporary research software development.

\subsection{Action Design Research (ADR)}\label{background1}

Action Design Research (ADR) is a design-oriented research methodology developed to support the creation and study of socio-technical artefacts in real-world contexts. Originally articulated by \citet{Sein2011}, ADR integrates elements of design science research and action research, emphasising the simultaneous generation of practical solutions and transferable design knowledge. Rather than treating artefact construction and evaluation as sequential or detached activities, ADR embeds design, use, and reflection within an iterative research process.

A central feature of ADR is its focus on Building–Intervention–Evaluation (BIE) cycles. In each cycle, artefacts are constructed, introduced within a relevant context, and examined in relation to both practical outcomes and research objectives. These cycles function as mechanisms for learning and abstraction rather than as endpoints in themselves. Reflection plays a critical role throughout the process, enabling researchers to move beyond local problem-solving toward more generalisable insights \citep{Sein2011}.

Subsequent elaborations of ADR have emphasised its adaptability across diverse research settings. \citet{Mullarkey2019} highlight that ADR can accommodate varying degrees of organisational involvement, artefact maturity, and evaluation formality, provided that iterative learning and reflective abstraction remain central. This flexibility has contributed to ADR’s uptake in information systems, educational technology, and digital innovation research, where artefacts evolve alongside stakeholder understanding.

ADR is particularly well-suited to research contexts in which artefacts function as instruments of inquiry rather than fixed endpoints. By legitimising iterative development and iterative evaluation, ADR allows design activity to serve as a vehicle for inquiry rather than as a purely technical means to an end. This orientation aligns closely with research software development practices, where understanding frequently emerges through implementation and experimentation rather than complete prior specification.

At the same time, ADR is intentionally abstract with respect to development practice. While it specifies the types of activities that should occur within a research cycle, it does not prescribe concrete mechanisms for organising day-to-day design work, managing development roles, or integrating emerging tools such as generative AI and AI agents. These aspects are left to contextual judgement, creating space for complementary frameworks that support consistent enactment in specific settings.

\subsection{Research software artefacts}\label{background2}

Research software refers to software artefacts developed primarily to support scholarly inquiry, experimentation, and knowledge generation rather than commercial deployment. Such software often embodies research assumptions, analytical models, or pedagogical intentions, and its design and behaviour can shape research outcomes. In contrast to production software, which is typically optimised for stability, scalability, and long-term maintenance, research software is frequently characterised by exploratory development and evolving requirements \citep{Wilson2014,Katz2021}.

Research software may take diverse architectural forms depending on the nature of the inquiry. In many contexts, artefacts consist of relatively lightweight implementations such as algorithms, analytical pipelines, simulation models, or computational representations used to explore theoretical or empirical questions. In other cases, particularly where interaction, data collection, or complex workflows are central, research software systems may take a more complex, multi-layered form. These full-stack configurations typically integrate user-facing interfaces, application logic, data persistence mechanisms, and execution or deployment environments. Even relatively small research projects may involve such end-to-end structures, requiring researchers to engage with system design beyond isolated scripts or analytical components.

The presence of full-stack research software has important methodological implications. Architectural decisions about data models, interaction flows, and deployment configurations can influence validity, reproducibility, interpretability, and the conditions under which evidence is generated. In such cases, development choices constitute a form of methodological decision-making rather than merely technical implementation. At the same time, similar epistemic considerations arise in non-full-stack artefacts, where algorithmic assumptions, modelling choices, or data-processing pipelines likewise shape research outcomes. Across this spectrum, the construction of research software is inseparable from the conduct of inquiry itself.

From an educational perspective, developing research software, whether lightweight or full-stack, can be an intensive learning process in which technical competence, domain understanding, and research reasoning co-evolve \citep{Papert1980,Kolb1984}. However, without structured guidance, development may become fragmented or overly driven by local technical concerns, undermining reflective learning and methodological coherence. These observations strengthen the case for practice-level scaffolding that connects research methodology with the realities of research software development, particularly under solo conditions where role multiplexing and AI-assisted work further complicate design and reflection.

\subsection{Solo research software development}\label{background3}

A defining characteristic of research software development in Higher Degree Research contexts is that it is most often undertaken by a single individual. Doctoral, master’s, and honours researchers frequently design, implement, evaluate, and maintain software artefacts independently, without the formal role separation, peer review structures, or governance mechanisms commonly found in industrial development environments.

Solo research software development entails a high degree of role multiplexing, in which a single researcher simultaneously acts as problem framer, domain expert, system architect, developer, tester, evaluator, and reflective analyst. These roles are epistemically consequential: decisions about system design and implementation directly shape the nature of the knowledge produced and the credibility of research claims. As a result, solo researchers must continually navigate between technical execution and methodological judgement.

While solo development offers advantages such as flexibility and tight integration between domain knowledge and system design, it also presents challenges. The absence of explicit role boundaries can obscure the rationale behind design decisions and complicate reflection on development activities. Cognitive load may become substantial as researchers manage multiple system layers alongside research planning, analysis, and documentation.

The increasing integration of generative AI tools further complicates solo development practice. As aspects of design, coding, or documentation are delegated to AI systems, boundaries between roles may become less visible. In the absence of explicit role differentiation and reflective practices, it can become unclear where human judgement ends and machine-generated contribution begins. Recognising solo research software development as a distinct context is therefore critical for understanding the methodological and learning challenges that arise in contemporary research practice.

\subsection{Human--AI collaboration in development work}\label{background4}

Human--AI collaboration has become an increasingly prominent paradigm in software development and other knowledge-intensive activities. Human-centred approaches frame AI systems as collaborators that augment human capabilities while preserving human responsibility for judgement, interpretation, and meaning-making \citep{Amershi2019,Shneiderman2020}. This framing is particularly relevant in research contexts, where transparency and accountability are essential.

In software development, AI-assisted tools now support activities such as code generation, refactoring, documentation, and architectural suggestions. These capabilities can reduce technical barriers for solo researchers and enable rapid exploration across system components. When used reflectively, such tools may function as cognitive scaffolds that support learning and experimentation. However, the benefits of AI assistance depend critically on how responsibilities and decision authority are distributed between humans and machines.

Research on human--AI interaction highlights the risks associated with poorly structured collaboration, including over-reliance on automation, reduced situational awareness, and diminished learning opportunities \citep{Amershi2019}. These risks are magnified in solo research settings, where researchers must independently determine how AI-generated contributions are interpreted, validated, and integrated into research artefacts.

Despite growing interest in Human–AI collaboration, existing literature offers limited guidance on how AI-assisted development practices should be aligned with structured research methodologies such as ADR. In particular, there is little articulation of how human judgement, reflective practice, and methodological coherence can be sustained when AI tools mediate substantial portions of development work. This gap motivates the need for clearer conceptual framing of Human–AI collaboration within research software development.

\section{Research Methodology}\label{metho}

This study adopts a pragmatic, design-oriented research paradigm in which knowledge is generated through the construction, use, and reflection on artefacts situated within research practice. Such a paradigm is well suited to contexts where understanding emerges through making and intervention rather than solely through observation or hypothesis testing. In line with this orientation, Action Design Research (ADR) is employed as the overarching research methodology guiding the study \citep{Sein2011}.

\subsection{Paradigm and methodological positioning}\label{metho1}

ADR is particularly appropriate for this research because it explicitly recognises socio-technical artefacts as central vehicles of inquiry. Rather than treating artefacts as static entities, ADR positions them as dynamic entities through which problems are explored and design knowledge is produced. Its emphasis on iterative Building–Intervention–Evaluation cycles aligns with the nature of research software development, where insight frequently arises through successive rounds of design and reflection.

Within the ADR tradition, this study focuses primarily on generating design knowledge rather than causal explanation. Following established guidance, the focus is on articulating transferable principles and structures that can inform similar research contexts beyond a single instantiation \citep{Mullarkey2019}. Accordingly, the contribution of this paper is conceptual rather than empirical, and evaluation is formative and reflective rather than summative.

A key methodological distinction made in this study is between research software artefacts and the framework that structures their creation. While research software systems may be produced through ADR cycles, they are not the unit of analysis in this paper. Instead, the SHAPR framework is treated as the primary design artefact and unit of analysis. This distinction avoids conflating framework contribution with system-specific implementation and clarifies the scope of the claims advanced.

\subsection{Unit of analysis and scope}\label{metho2}

The primary unit of analysis in this study is the SHAPR framework itself. SHAPR is examined as a design artefact whose purpose is to operationalise ADR for solo, AI-assisted research software development. Analysis therefore focuses on the framework’s internal coherence, conceptual completeness, and alignment with ADR principles rather than on the performance or characteristics of any specific software system.

Research software artefacts occupy a contextual role within this study. They represent the types of systems that may be developed when SHAPR is enacted, but they are not evaluated or analysed here. This scoping decision positions the paper as a foundational contribution within a broader research programme, establishing a coherent framework prior to empirical application.

\subsection{Evaluation approach}\label{metho3}

Evaluation in ADR is understood as an ongoing and formative activity embedded within iterative cycles of design and reflection. Consistent with this view, the evaluation adopted in this study is reflective and analytical rather than empirical. The SHAPR framework is examined in terms of its alignment with ADR principles, its suitability for solo research contexts, and its capacity to support human accountability and learning in AI-assisted development environments.

This form of evaluation prioritises analytical generalisation through the articulation of design knowledge rather than empirical validation in a specific domain \citep{Mullarkey2019}. The intent is to establish SHAPR as a coherent and defensible framework that warrants subsequent empirical exploration rather than as a finalised solution.

\section{The SHAPR Framework}\label{shapr}

This section presents the SHAPR framework as the central contribution of the study. Building on the methodological foundation established in the preceding sections, SHAPR is introduced as a practice-level framework that operationalises Action Design Research (ADR) for solo, Human-centred and AI-assisted research software development. The section positions SHAPR conceptually, articulates its design principles, and outlines its structure. It then explains how role multiplexing, artefact and evidence management, and the tool ecosystem are integrated to support rigorous research practice and learning.

\subsection{Conceptual positioning of SHAPR}\label{shapr1}

SHAPR is proposed as a practice-level operational framework to support the enactment of ADR in the specific context of solo, Human-centred and AI-assisted research software development. SHAPR---\emph{Solo, Human-centred, AI-assisted PRactice}---does not introduce a new research methodology nor alter ADR’s epistemological commitments. Rather, it provides an operational scaffold that translates ADR’s high-level principles into actionable guidance for contemporary research software practice \citep{Sein2011,Mullarkey2019}.

\begin{figure}[h]
\centering
\includegraphics[width=0.9\textwidth]{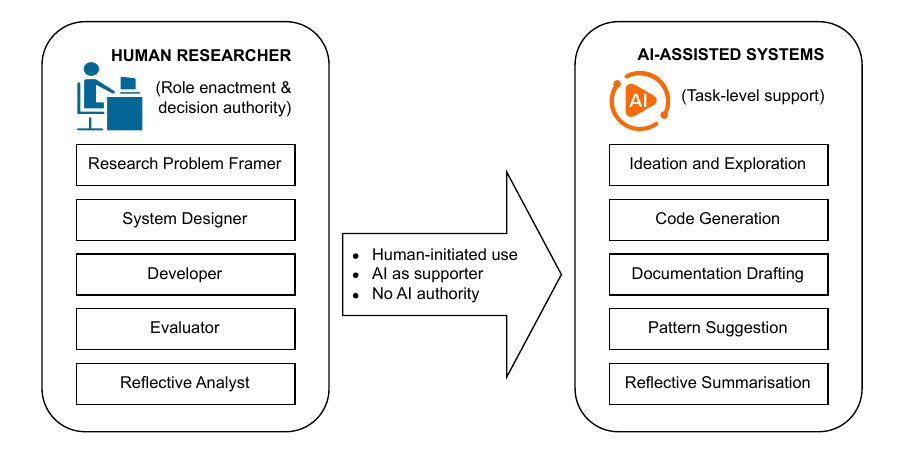}
\caption{Human--AI role multiplexing in solo research software development under SHAPR. The figure depicts conceptual human roles enacted by a solo researcher alongside AI-assisted support functions, emphasising human-centred governance and the non-delegation of decision authority to AI systems.}
\label{fig:shapr_roles}
\end{figure}

In this study’s conceptual architecture, ADR functions as the overarching research methodology: it defines the logic of inquiry, the role of artefacts, and the iterative Building--Intervention--Evaluation (BIE) cycles through which design knowledge is generated \citep{Sein2011}. SHAPR operates at a lower level of abstraction, focusing on how BIE cycles may be enacted when research software is developed by a single researcher working with generative AI tools. Figure~\ref{fig:shapr_roles} (and Figure~\ref{fig:shapr_overview}) locate SHAPR as an operational layer between ADR and the evolving research software artefacts that serve as vehicles for inquiry, consistent with a design knowledge perspective on artefacts and guidance \citep{Hevner2004,GregorHevner2013}.

A central premise is that solo research software development constitutes a distinctive socio-technical setting in which one person must enact multiple roles typically distributed across teams, including problem framing, system design, implementation, evaluation, and reflection. SHAPR addresses this condition by emphasising \emph{conceptual} role separation rather than physical role distribution. The framework encourages the researcher to make explicit which role is being enacted at a given moment (e.g., designer, developer, evaluator, reflective analyst), thereby supporting methodological clarity, accountability, and learning. This role multiplexing logic is summarised in Table~\ref{tab:shapr_roles} and discussed further in Section~\ref{shapr4}.

Human-centred governance is a defining characteristic of SHAPR. While the framework acknowledges the growing role of generative AI in research software development, it places human judgement, accountability, and sense-making at the centre of all activities. AI systems are treated as \emph{cognitive collaborators} that may support ideation, implementation, and reflection, but they are not framed as autonomous agents or decision-makers. This positioning is consistent with established principles in human-centred AI and human--AI interaction, and is intended to preserve epistemic responsibility with the human researcher \citep{Amershi2019,Shneiderman2020}. The allocation of responsibilities between the human researcher and AI-assisted systems is summarised in Table~\ref{tab:shapr_human_ai}.

SHAPR is intentionally tool-agnostic. Although the paper illustrates how representative tool categories may support SHAPR enactment, the framework is not tied to specific platforms or vendors. This design choice supports transferability across domains and institutional contexts, and improves resilience to rapid changes in tool ecosystems. Table~\ref{tab:shapr_tools} provides illustrative mappings between tool categories and SHAPR activities; these examples demonstrate applicability rather than exhaustiveness and may be substituted as new tools emerge.

Overall, SHAPR contributes practice-level guidance consistent with the design knowledge tradition underlying ADR. It clarifies how methodological intent may be realised in day-to-day development without prescribing specific solutions or constraining researcher creativity. In doing so, SHAPR functions both as a scaffold for rigorous research practice and as a learning support for solo researchers navigating AI-assisted software development \citep{GregorHevner2013,Schon1983,Kolb1984}.

\subsection{Design principles underpinning SHAPR}\label{shapr2}

SHAPR is grounded in explicit design principles that translate ADR’s commitments into practical guidance for solo, Human-centred and AI-assisted research software development. The principles were derived through reflective synthesis of methodological requirements of ADR and recurring challenges in solo research practice and Human--AI collaboration \citep{Sein2011,Mullarkey2019}. They articulate normative assumptions that shape how SHAPR is intended to be enacted and how its use may be evaluated.

First, SHAPR prioritises \emph{human accountability over automation}. Regardless of the degree of AI assistance, responsibility for research decisions, interpretations, attribution, and outcomes remains with the human researcher. AI systems may support ideation, implementation, or documentation, but do not replace judgement or epistemic responsibility \citep{Amershi2019,Shneiderman2020}.

Second, SHAPR emphasises \emph{conceptual role separation in solo practice}. Because solo researchers enact multiple roles, SHAPR encourages explicit cognitive separation of design, development, evaluation, and reflection to support clearer reasoning and reduce implicit role drift. This principle aligns with ADR’s emphasis on reflection and abstraction as mechanisms for producing transferable design knowledge \citep{Sein2011,Schon1983}. Role mappings and responsibilities are illustrated in Table~\ref{tab:shapr_roles}.

Third, SHAPR promotes \emph{iterative learning through artefact-centred reflection}. Consistent with ADR’s BIE cycles, SHAPR treats research software artefacts as evolving objects through which learning occurs. Reflection is embedded throughout development rather than deferred to the end, reducing the risk of retrospective rationalisation and strengthening the link between design action and research reasoning \citep{Sein2011,Kolb1984}.

Fourth, SHAPR requires \emph{explicit artefact and version traceability}. Versioned software components, configurations, prompts, and reflective notes are treated as integral research evidence rather than ancillary documentation. This supports transparency and enables reconstruction of design trajectories and learning across iterations, consistent with expectations for rigour in design-oriented research \citep{GregorHevner2013,Hevner2004}. Table~\ref{tab:shapr_artefacts} summarises expected artefacts and evidence types.

Fifth, SHAPR adopts \emph{tool agnosticism with contextual sensitivity}. The framework accommodates diverse tools that may support development, reflection, and evidence practices, recognising that tool ecosystems evolve rapidly. Illustrative mappings between tool categories and framework activities are provided in Table~\ref{tab:shapr_tools} to clarify practical enactment while avoiding dependence on particular technologies.

Finally, SHAPR is underpinned by \emph{learning-oriented governance}. The framework promotes lightweight governance mechanisms---such as reflective checkpoints and role-aware reviews---that support learning, accountability, and methodological coherence without constraining exploration. This principle aligns SHAPR with educational aims in research training and technology-supported learning \citep{Schon1983,Kolb1984}.

Taken together, these principles define the conceptual boundaries of SHAPR and provide criteria against which the framework can be refined. Table~\ref{tab:shapr_principles} summarises the principles and their implications for solo, AI-assisted research software development.

\begin{table}[h]
\caption{Design principles underpinning the SHAPR framework}\label{tab:shapr_principles}
\begin{tabular}{@{}p{2.5cm}p{5cm}p{6cm}@{}}
\toprule
\textbf{Design Principle} & \textbf{Description} & \textbf{Rationale and Implications} \\
\midrule
Human-centred governance &
Decision authority, accountability, and reflective judgement remain with the human researcher throughout the development process. &
Ensures methodological and ethical responsibility in AI-assisted development. Aligns with ADR’s emphasis on reflection and prevents delegation of epistemic responsibility to AI systems. \\

Operationalisation of ADR in practice &
Provides practice-level guidance for enacting Action Design Research through iterative Building--Intervention--Evaluation (BIE) cycles. &
Bridges the gap between high-level ADR methodology and day-to-day research software development, particularly for solo researchers. \\

Role multiplexing with conceptual separation &
Recognises that a solo researcher enacts multiple roles while encouraging explicit conceptual separation between them. &
Supports reflective awareness, reduces cognitive overload, and clarifies Human--AI collaboration by separating role responsibility from task-level assistance. \\

AI as cognitive collaborator, not decision-maker &
Positions generative AI tools as supporting ideation, coding, documentation, and pattern recognition without assuming decision authority. &
Preserves human accountability while enabling productive AI assistance; aligns with ethical and pedagogical concerns in AI-supported research and learning. \\

Artefact-centred evidence and traceability &
Treats research software artefacts, documentation, prompts, planning notes, and reflections as evolving research evidence. &
Supports transparency, traceability, and analytical generalisation across ADR cycles by linking artefacts to decisions and learning outcomes. \\

Lightweight but rigorous documentation &
Promotes intentional documentation practices that capture reasoning and learning without imposing excessive administrative burden. &
Balances methodological rigour with the practical constraints of solo research, ensuring documentation supports reflection rather than impeding progress. \\

Tool-agnostic enablement &
Focuses on roles, practices, and artefacts rather than prescribing specific tools or platforms. &
Ensures transferability and longevity of the framework across evolving AI and development ecosystems. \\

Learning-oriented design &
Frames research software development as a learning process supporting reflection, abstraction, and skill development. &
Aligns the framework with educational research goals and the focus of \emph{Technology, Knowledge and Learning} on technology-supported learning. \\
\bottomrule
\end{tabular}
\end{table}

\subsection{Framework structure and overview}\label{shapr3}

\begin{figure}[h]
\centering
\includegraphics[width=0.9\textwidth]{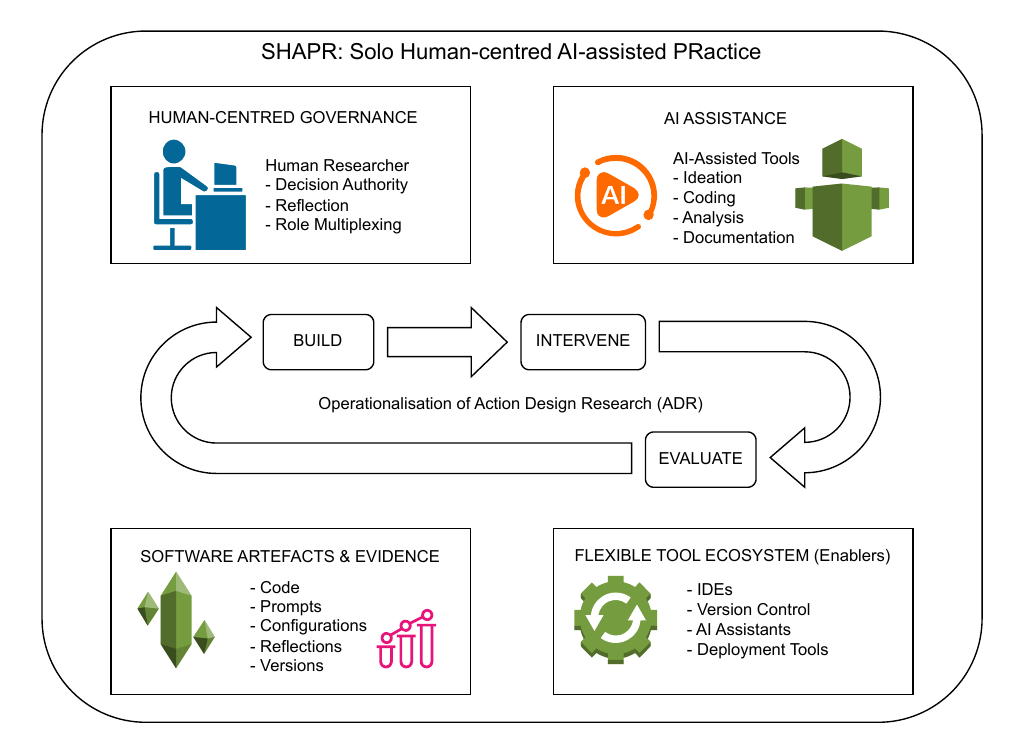}
\caption{Overview of the SHAPR framework. The figure illustrates the conceptual structure of SHAPR, positioning Action Design Research as the guiding methodology for solo research software development. The framework integrates iterative Building--Intervention--Evaluation cycles, human--AI collaboration, and artefact-centred evidence generation to support reflective and learning-oriented research practice.}
\label{fig:shapr_overview}
\end{figure}

SHAPR provides an operational scaffold for enacting ADR in solo, Human-centred and AI-assisted research software development. Rather than prescribing a linear workflow, it organises practice around iterative BIE cycles and makes explicit how these cycles are enacted through role awareness, artefact management, and reflective checkpoints \citep{Sein2011,Mullarkey2019}. Figure~\ref{fig:shapr_overview} presents an overview of the framework.

At a high level, SHAPR can be understood as three interrelated layers: (1) ADR as the methodological foundation that defines epistemic intent and iterative BIE logic; (2) SHAPR as the operational layer that specifies practices for enacting ADR in solo contexts; and (3) research software artefacts as evolving outcomes that function as vehicles for inquiry rather than endpoints. This layered relationship aligns with design science perspectives that treat artefacts and associated guidance as central to research contribution \citep{GregorHevner2013,Hevner2004}.

Within the operational layer, activities align with BIE cycles: \emph{Building} includes requirements articulation, architectural exploration, implementation, and revision of artefacts (often supported by AI-assisted tools); \emph{Intervention} involves using the artefact in its intended research or learning context; and \emph{Evaluation} encompasses technical assessment and reflective analysis of decisions, behaviours, and AI-assisted contributions. Reflection and abstraction are embedded across activities rather than isolated as a final phase, consistent with ADR’s iterative learning orientation \citep{Sein2011}.

A defining feature of SHAPR is explicit visibility of roles and artefacts. The framework encourages the researcher to externalise artefacts (code versions, prompts, design notes, reflections) and to acknowledge role transitions across cycles. This visibility supports traceability and learning by making the evolution of artefacts and reasoning inspectable over time. Expected artefacts and evidence types are summarised in Table~\ref{tab:shapr_artefacts}.

SHAPR also incorporates lightweight governance mechanisms as reflective checkpoints within and between BIE cycles. These are not formal approval stages, but deliberate pauses to review artefact changes, assess the influence of AI assistance, and check alignment with research aims. Figure~\ref{fig:shapr_evidence} illustrates how artefacts and evidence accumulate across cycles to support reflection, abstraction, and reporting.

Overall, SHAPR provides a structured yet adaptable approach for solo researchers enacting ADR in AI-assisted software development. By making methodological intent, development activity, and reflective practice visible within a unified structure, SHAPR supports rigorous research and meaningful learning. The following sections elaborate role multiplexing (Section~4.4), artefact and evidence management (Section~4.5), and tool ecosystem enablement (Section~4.6).

\subsection{Human--AI role multiplexing in solo contexts}\label{shapr4}

A central challenge addressed by SHAPR is role multiplexing in solo research software development. Whereas team-based settings distribute responsibilities across specialised roles, solo researchers must enact multiple roles throughout the research process, including problem framing, system design, development, evaluation, and reflection. Without explicit differentiation, role drift can obscure reasoning, accountability, and learning.

SHAPR responds by promoting conceptual role separation. While all roles may be enacted by one person, the framework encourages the researcher to explicitly identify which role is being enacted and to reflect on how transitions influence design decisions and research outcomes. This role logic aligns with ADR’s emphasis on reflection and abstraction as mechanisms for transforming situated design activity into transferable knowledge \citep{Sein2011,Schon1983}. Table~\ref{tab:shapr_roles} summarises roles and responsibilities.

Generative AI further complicates role boundaries because AI tools may contribute to activities typically associated with human roles, such as generating code, proposing architectural patterns, or drafting documentation. SHAPR therefore treats AI systems as role-supporting collaborators rather than role holders: AI may assist with tasks within a role, but responsibility for the role remains with the human researcher. The allocation of responsibilities is summarised in Table~\ref{tab:shapr_human_ai}.

Role awareness also supports learning. By making role transitions explicit and recording how AI assistance influenced decisions, SHAPR encourages reflection not only on what was produced, but on how design reasoning evolved. This reflective visibility supports development of transferable skills in critical thinking, design reasoning, and responsible AI use \citep{Amershi2019,Schon1983}.

SHAPR does not prescribe rigid sequences or formalised role transition checkpoints. Instead, it provides a conceptual vocabulary and structure that researchers can adapt to their projects and constraints while maintaining methodological coherence. The next section explains how artefact and evidence practices support this role-aware enactment.

\begin{table}[h]
\caption{Human roles and responsibilities in solo research software development under SHAPR}\label{tab:shapr_roles}
\begin{tabular}{@{}p{1.5cm}p{4.3cm}p{4.5cm}p{4.0cm}@{}}
\toprule
\textbf{Role} & \textbf{Primary responsibilities} & \textbf{Illustrative activities} & \textbf{Relationship to ADR} \\
\midrule
Research Problem Framer &
Interprets and bounds the research problem; articulates research intent; defines scope, assumptions, and constraints. &
Clarifying objectives and constraints; formulating research questions; producing and revising problem statements and planning notes. &
Supports problem formulation and abductive reasoning that initiates and guides ADR cycles. \\

System Designer &
Translates research intent into conceptual and architectural designs aligned with inquiry and experimentation. &
Defining architecture; exploring alternative structures; analysing trade-offs; documenting design rationale. &
Shapes design interventions by aligning system structure with research aims. \\

Developer &
Implements and modifies research software artefacts in accordance with design decisions and emerging insights. &
Writing and refactoring code; integrating components; configuring environments; experimenting with implementation alternatives. &
Enacts the building component of ADR through iterative artefact construction and refinement. \\

Evaluator &
Assesses artefacts and development outcomes against research objectives, technical expectations, and methodological criteria. &
Testing functionality; assessing fitness-for-purpose; identifying unexpected behaviours and limitations; recording evaluation evidence. &
Contributes to evaluation by examining the effects of interventions and informing refinement. \\

Reflective Analyst &
Synthesises insights from development activities and evaluation outcomes to support learning, abstraction, and reporting. &
Documenting reflections; analysing decisions; identifying recurring patterns; articulating lessons learned and abstractions. &
Enables reflection and abstraction, supporting knowledge generation across ADR iterations. \\
\bottomrule
\end{tabular}
\end{table}

\begin{table}[h]
\caption{Allocation of responsibilities between the human researcher and AI-assisted systems in SHAPR}\label{tab:shapr_human_ai}
\begin{tabular}{@{}p{2cm}p{5.0cm}p{5.0cm}@{}}
\toprule
\textbf{Aspect} & \textbf{Human researcher responsibilities} & \textbf{AI-assisted system support} \\
\midrule
Problem interpretation and framing &
Interprets the research context; defines and refines the research problem; determines scope, assumptions, and boundaries; ensures alignment with research objectives. &
Generates alternative framings, questions, or perspectives to support exploratory sense-making under human direction. \\

Design decision-making &
Selects and justifies architectural and design approaches; evaluates trade-offs; ensures coherence with research aims and theoretical grounding. &
Suggests candidate architectural or design patterns based on analogous prior solutions to prompt human consideration. \\

Implementation choices &
Determines what to implement, modify, or discard; integrates components; ensures correctness and suitability of code within the research context. &
Generates candidate code snippets, configurations, or implementation alternatives to accelerate development. \\

Evaluation and judgement &
Evaluates artefacts against research objectives and methodological criteria; interprets outcomes; identifies limitations and implications. &
Assists with summarisation of evaluation outcomes, identification of anomalies, or articulation of evaluation criteria for human review. \\

Reflection and abstraction &
Conducts reflective analysis across iterations; synthesises insights; abstracts transferable knowledge and lessons learned. &
Supports reflective summarisation by highlighting patterns, contrasts, or recurring themes across artefacts and documentation. \\

Accountability and ethics &
Retains full responsibility for research integrity, ethical compliance, attribution, and reporting. &
Does not assume responsibility or accountability; operates solely as a supporting tool under human governance. \\
\bottomrule
\end{tabular}
\end{table}

\subsection{Artefact, versioning, and evidence management}\label{shapr5}

Rigorous research practice requires traceability of how insights, decisions, and outcomes emerge over time. In research software development, this includes intermediate versions, rationales, prompts, and reflective records in addition to the final software artefact. SHAPR therefore foregrounds artefact, versioning, and evidence management as integral research practices rather than ancillary documentation.

Within SHAPR, research software artefacts are treated as evolving objects of inquiry. Each iteration reflects design decisions, assumptions, and interactions with AI-assisted tools that contribute to understanding of the problem space. Maintaining visibility of change across BIE cycles supports the reflective and abductive reasoning central to ADR \citep{Sein2011,Mullarkey2019}.

Versioning provides a practical mechanism for this visibility. SHAPR encourages systematic versioning of software components, configuration files, prompts, and supporting documentation so that researchers can reconstruct design trajectories and relate changes to emerging research insights. Importantly, versioning is framed as a methodological support mechanism for transparency and analytical generalisation, not merely an imported engineering convention \citep{GregorHevner2013,Hevner2004}. Table~\ref{tab:shapr_artefacts} summarises expected artefacts and evidence types.

In solo contexts, evidence practices also serve a governance function. Without collaborators to provide informal checks or shared memory, researchers risk losing sight of why design choices were made or how AI assistance influenced outcomes. SHAPR addresses this risk through lightweight documentation practices, such as reflective notes linked to specific artefact versions and role-aware annotations that record when AI assistance was used.

\begin{figure}[h]
\centering
\includegraphics[width=0.95\textwidth]{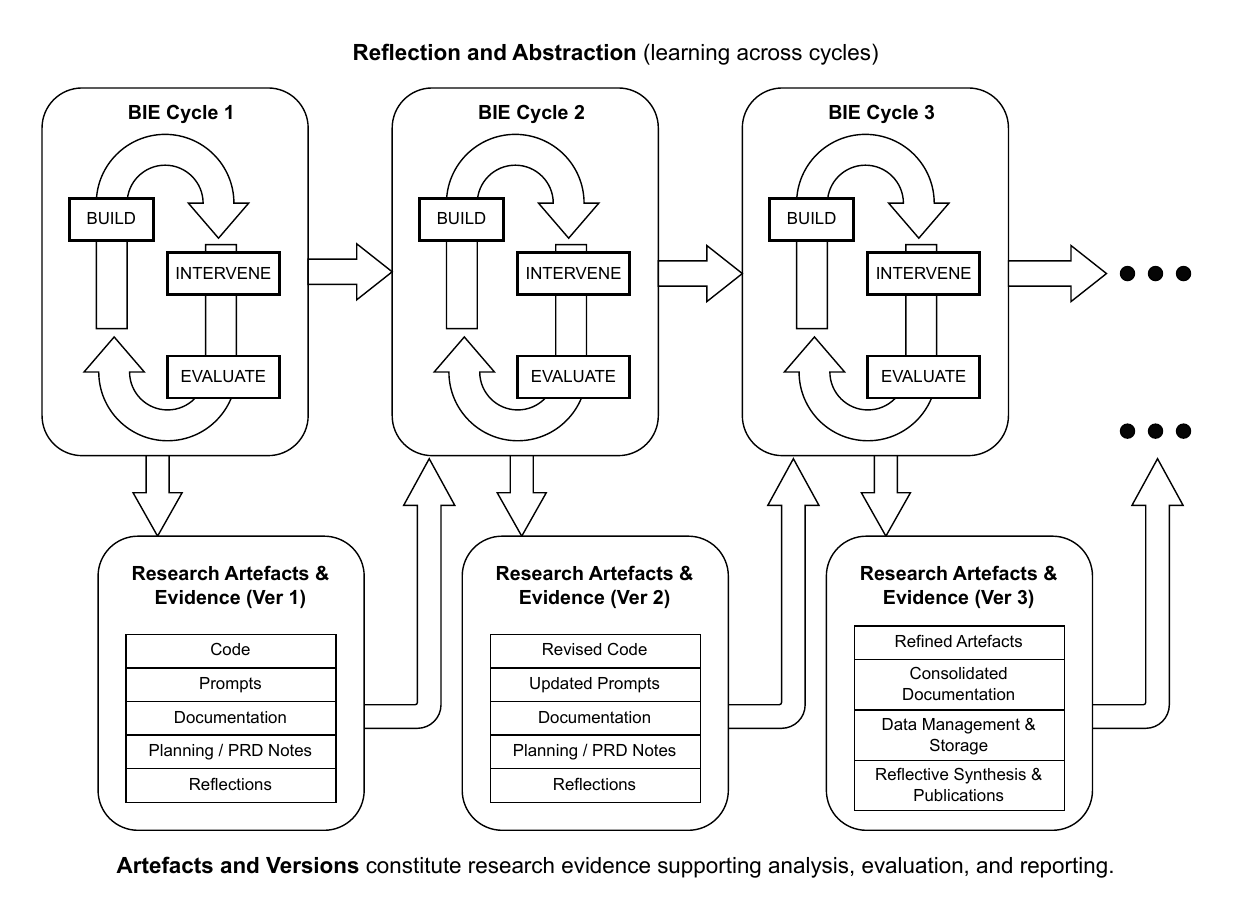}
\caption{Artefact and evidence flow across Action Design Research Building--Intervention--Evaluation cycles. The figure shows how research software artefacts, documentation, prompts, and reflections evolve iteratively and are treated as research evidence supporting reflection, abstraction, and learning across cycles.}
\label{fig:shapr_evidence}
\end{figure}

Artefact-centred evidence management also supports learning. By examining how artefacts and decisions evolved, researchers can identify patterns, challenges, and transferable insights beyond a single project. Figure~\ref{fig:shapr_evidence} illustrates evidence accumulation across cycles.

\begin{table}[h]
\caption{Research artefacts and evidence types in the SHAPR framework}\label{tab:shapr_artefacts}
\begin{tabular}{@{}p{2cm}p{4cm}p{4cm}p{4cm}@{}}
\toprule
\textbf{Artefact / evidence type} & \textbf{Description} & \textbf{Role in the research process} & \textbf{Relationship to ADR} \\
\midrule
Source code &
Executable code implementing research software functionality, including prototypes and experimental features. &
Material instantiation of design interventions; evolves across iterations to reflect emerging insights. &
Constitutes the building component of ADR and provides a basis for evaluation and reflection. \\

Configuration and environment files &
Settings, dependencies, and environment specifications required to execute or reproduce the software artefact. &
Supports transparency and reproducibility; documents contextual assumptions underlying development and evaluation. &
Enables rigorous evaluation by making interventions inspectable and repeatable. \\

Prompts and AI interaction records &
Prompts, instructions, and interaction traces generated when engaging AI-assisted systems. &
Provide visibility into how AI assistance influenced development activities and decisions. &
Support reflective analysis of Human--AI collaboration across ADR cycles. \\

Planning and PRD documentation &
Evolving notes capturing intent, scope, assumptions, and design rationale, including informal or lightweight PRD artefacts. &
Articulates research intent and design direction; revised iteratively as understanding develops. &
Supports problem framing and abductive reasoning that inform successive interventions. \\

Design sketches and architectural notes &
Conceptual diagrams, structural descriptions, or design annotations produced during system design. &
Externalise design reasoning and enable comparison of alternative structures and decisions. &
Shape intervention design and support abstraction across iterations. \\

Evaluation notes and test records &
Observations, test outcomes, and assessment notes produced during evaluation activities. &
Capture evidence of artefact performance relative to research objectives and expectations. &
Provide empirical grounding for the evaluation phase of ADR. \\

Reflective journals and analytic memos &
Written reflections capturing insights, challenges, decisions, and lessons learned across cycles. &
Support learning, sense-making, and articulation of transferable knowledge. &
Enable reflection and abstraction central to ADR knowledge generation. \\

Version history and change logs &
Records of artefact evolution across iterations, including commits, revisions, or version annotations. &
Provide traceability of design decisions and changes over time. &
Link artefacts to specific ADR cycles, supporting transparency and analytical generalisation. \\
\bottomrule
\end{tabular}
\end{table}

\subsection{Tool ecosystem as enablers}\label{shapr6}

Research software development is increasingly mediated by a diverse ecosystem of tools, including generative AI systems, AI-assisted development environments, version control platforms, and deployment infrastructures. Within SHAPR, such tools are positioned as enablers of practice rather than determinants of methodology. This distinction preserves a human-centred, methodologically grounded orientation while acknowledging practical realities of AI-assisted development.

SHAPR remains tool-agnostic at the conceptual level: it specifies roles, activities, and artefacts that support ADR enactment without prescribing vendors or workflows. This improves transferability and resilience to ongoing changes in tool ecosystems. Table~\ref{tab:shapr_tools} provides illustrative mappings between representative tool categories and SHAPR activities; the examples demonstrate applicability rather than exhaustiveness and may be substituted as new tools emerge.

Generative AI systems play a prominent role in this ecosystem. Within SHAPR, conversational assistants may support problem exploration, requirements articulation, prompt-based prototyping, and reflective summarisation. They may also support design reasoning by suggesting candidate architectural or design patterns based on analogous prior solutions. For example, AI systems may assist with drafting preliminary requirements, generating candidate code during Building activities, and synthesising reflective insights during Evaluation phases. Importantly, these tools augment human reasoning rather than replace it; responsibility for interpreting, validating, and integrating AI-generated outputs remains with the human researcher \citep{Amershi2019,Shneiderman2020}
.

Vibe coding and AI-assisted development environments can accelerate exploratory development by translating high-level intentions into early prototypes. In SHAPR, such acceleration is most appropriate in early or exploratory BIE cycles where iteration supports learning and sense-making. The framework’s emphasis on traceability and reflective checkpoints is intended to ensure that speed does not come at the expense of methodological transparency.

Version control platforms support artefact and evidence management by maintaining change history, linking revisions to documentation, and enabling reconstruction of design trajectories across cycles. In SHAPR, these capabilities underpin methodological practices rather than serving purely technical ends, reinforcing versioning as a research support mechanism.

Overall, SHAPR conceptualises the tool ecosystem as a flexible set of resources that can be selectively appropriated to support different aspects of research software development. By decoupling methodological intent from specific technologies, the framework accommodates emerging tools while maintaining a stable, human-centred core.

\begin{table}[h]
\caption{Illustrative mapping between representative tool categories and SHAPR activities}\label{tab:shapr_tools}
\begin{tabular}{@{}p{2cm}p{2.5cm}p{4cm}p{4.8cm}@{}}
\toprule
\textbf{Tool category} & \textbf{Illustrative examples} & \textbf{Supported SHAPR activities} & \textbf{Nature of support (illustrative)} \\
\midrule
Conversational AI assistants &
ChatGPT; Gemini; Claude (or equivalents) &
Problem exploration; requirements articulation; reflective summarisation; evaluation synthesis &
Generates alternative framings and candidate text; supports reflective summarisation; helps articulate assumptions and rationale under human governance. \\

Vibe coding / rapid prototyping tools &
Emergent; Replit AI; Lovable; Bolt; v0; Google Anti-Gravity (or equivalents) &
Prompt-based prototyping; rapid implementation scaffolding; early iteration planning within Build activities &
Accelerates creation of exploratory prototypes and full-stack scaffolds from high-level intent; supports fast learning-oriented iteration while maintaining human decision authority. \\

AI-assisted IDEs and coding copilots &
GitHub Copilot; Cursor; JetBrains AI; Amazon CodeWhisperer; Tabnine; Codeium (or equivalents) &
Code generation; refactoring; pattern suggestion; documentation drafting within Build activities &
Supports localised coding tasks, refactoring, and pattern exploration; reduces routine effort while requiring human validation and integration. \\

Version control and repository platforms &
GitHub; GitLab; Bitbucket (or equivalents) &
Versioning; traceability; evidence capture; collaboration (where applicable) &
Maintains change history as research evidence; supports reproducibility and structured iteration across ADR cycles. \\

Issue tracking and lightweight planning tools &
GitHub Issues; Jira; Trello (or equivalents) &
Planning artefacts; task tracking; PRD notes; iteration coordination &
Externalises intent and scope; captures decisions and evolving requirements as versioned evidence. \\

Documentation and knowledge management tools &
Markdown; Overleaf/LaTeX; wikis; Notion (or equivalents) &
Documentation; reflective journaling; reporting; artefact interpretation &
Captures design rationale, reflections, and learnings; supports abstraction and reporting from evidence. \\

Testing, evaluation, and analysis tooling &
Unit test frameworks; notebooks; logging/monitoring tools (or equivalents) &
Evaluation records; validation; performance checks; evidence generation &
Supports systematic evaluation and recording of outcomes; helps produce inspectable evaluation evidence. \\

Deployment and environment tooling &
Containers; virtual environments; CI/CD pipelines (or equivalents) &
Reproducible builds; deployment; environment specification; intervention packaging &
Supports repeatability and portability of interventions; enables evaluation in controlled environments. \\
\bottomrule
\end{tabular}
\end{table}

\section{Scope, Applicability, and Research Programme Positioning}\label{scope}

This section clarifies the intended scope and boundaries of SHAPR and positions the present study within a broader Action Design Research (ADR) programme. It specifies the contexts for which SHAPR is designed---solo research software development in educational and scholarly settings---and explains how the conceptual contribution in this paper provides a foundation for subsequent empirical applications. This positioning reinforces the framework-focused nature of the present study while establishing clear pathways for future research \citep{Sein2011,Mullarkey2019,GregorHevner2013}.

\subsection{Intended scope of SHAPR}\label{scope1}

SHAPR is scoped to support solo researchers developing research artefacts in educational and scholarly contexts. It is designed primarily for Higher Degree Research (HDR) environments, including doctoral, master’s, and honours projects, where software-based artefacts are frequently developed as part of the research method and where learning, reflection, and methodological rigour are central concerns \citep{Schon1983,Kolb1984}. Within this scope, SHAPR responds to practical conditions of solo development, artefact complexity (including but not limited to full-stack systems), and the growing use of generative AI tools.

SHAPR is not intended as a general-purpose software engineering methodology, nor as a replacement for established processes used in industrial or large-team settings. Its design reflects constraints and affordances typical of solo research practice: limited resources, role multiplexing, and the dual obligation to produce functional artefacts and scholarly outputs. While elements of SHAPR may be informative in other settings, its primary contribution is support for research-oriented development rather than production-grade engineering.

The framework is further scoped to research software that functions as a vehicle for inquiry, experimentation, or learning---for example, systems built to explore research questions, simulate phenomena, collect or analyse data, or support reflective investigation. SHAPR does not assume that such software will be deployed at scale, maintained over long periods, or optimised for commercial use. Instead, it prioritises methodological alignment, traceability, and learning outcomes, consistent with design-oriented research in which artefacts and associated practices constitute core evidence and contribution \citep{Hevner2004,GregorHevner2013}.

With respect to AI use, SHAPR is intended for contexts in which generative AI tools are employed as development aids rather than autonomous agents. AI assistance is treated as augmenting human reasoning and reducing development friction, while decision authority and epistemic responsibility remain with the human researcher \citep{Amershi2019,Shneiderman2020}. Accordingly, SHAPR does not address settings where software is generated and evolved with minimal human involvement, nor does it aim to govern fully autonomous agentic development workflows.

From an educational perspective, SHAPR is particularly relevant where research software development is also a learning process---including research training, capstone-style projects, and practice-based research in which reflection on development activities contributes to knowledge formation \citep{Schon1983}. By embedding reflective checkpoints and artefact-centred evidence practices, SHAPR aims to support both research outcomes and researcher training and development.

By delimiting its intended scope, SHAPR avoids claims beyond the evidence presented in this paper while remaining transferable in principle across research domains. The framework is designed to be stable at the conceptual level yet adaptable in practice, providing a methodological scaffold that can be tailored to specific contexts. The following subsection positions this conceptual contribution within a broader research programme and clarifies how SHAPR may be applied and evaluated empirically.

\subsection{Positioning within a broader research programme}\label{scope2}

The contribution of this paper is intentionally positioned as a foundational stage within a broader Action Design Research (ADR) programme. Rather than reporting a single system instantiation, the study focuses on articulating and justifying SHAPR as a coherent and transferable form of design knowledge that operationalises ADR for solo, Human-centred and AI-assisted, research software development. This positioning reflects established guidance in design-oriented research, where frameworks and principles may constitute standalone contributions that enable and structure subsequent empirical work \citep{Sein2011,GregorHevner2013,Mullarkey2019}.

In this study, the SHAPR framework itself is treated as the primary design artefact and unit of analysis. The evaluation undertaken is formative and reflective, focusing on internal coherence, conceptual completeness, alignment with ADR principles, and suitability for the distinctive characteristics of solo research practice. This approach is consistent with ADR’s emphasis on generating reusable design knowledge through abstraction and reflection, rather than limiting contribution to the performance of a single artefact instance \citep{Sein2011}.

Once articulated, SHAPR is intended to be applied across multiple research software development projects rather than tied to a single case. The framework may be enacted in a wide range of Higher Degree Research (HDR) contexts, including doctoral, master’s, and honours theses, where software artefacts are developed to support inquiry, experimentation, or learning. These applications may span diverse disciplinary domains, such as engineering, information systems, data analytics, cybersecurity, educational technology, and other computationally mediated research fields.

In such applications, research software artefacts and their development trajectories may become the focal objects of empirical investigation. SHAPR provides a stable methodological scaffold that enables researchers to examine how Human--AI collaboration, role multiplexing, artefact traceability, and reflective practices shape both research outcomes and learning processes. Insights generated through these applications can, in turn, inform refinement of the framework, supporting further cycles of reflection and abstraction consistent with the iterative logic of ADR \citep{Mullarkey2019}.

By positioning SHAPR as a reusable and extensible framework rather than a single-use method, this study supports cumulative knowledge building across research projects and contexts. Following well-established research principles and practices \citep{Bhatt2012}, the framework is designed to remain conceptually stable while being adaptable in practice, allowing it to accommodate evolving tools, research questions, and institutional settings. This programme-oriented positioning ensures that SHAPR functions not only as guidance for individual projects, but also as an integrative structure for sustained research and learning in AI-assisted research software development.

\section{Discussion}\label{discuss}

This section discusses the implications of the SHAPR framework for research practice and learning in software-intensive research contexts. Building on the conceptual contribution established in the preceding sections, it examines how SHAPR informs solo research software development, the responsible integration of generative AI into research workflows, and the design of learning and supervision practices. The section concludes by reflecting on the limitations of the present study, situating SHAPR as a foundational contribution within an ongoing design-oriented research agenda rather than a final or exhaustive solution.

\subsection{Implications for research practice}\label{discuss1}

The SHAPR framework has several implications for research practice in contexts where software artefacts are central to inquiry and development is undertaken by a solo researcher. By operationalising Action Design Research (ADR) for AI-assisted software development, SHAPR provides a structured means of integrating software construction into methodological reasoning rather than treating it as a peripheral technical activity \citep{Sein2011,GregorHevner2013}.

A primary implication concerns methodological rigour. SHAPR foregrounds artefact traceability, role awareness, and iterative reflection, enabling researchers to make explicit connections between software design decisions, research questions, and emerging insights. This visibility is particularly important in Higher Degree Research (HDR) contexts, where examiners and reviewers often scrutinise the relationship between methodological claims and technical artefacts. By providing a conceptual scaffold for articulating this relationship, SHAPR supports more defensible and transparent research practice \citep{Hevner2004}.

SHAPR also addresses the cognitive and organisational challenges of solo research practice. Solo researchers routinely shift between conceptual reasoning, technical implementation, and evaluative reflection, often under conditions of limited time and support. By encouraging conceptual role separation, SHAPR provides a vocabulary and structure that helps researchers manage these transitions explicitly, reducing the risk that methodological considerations are overshadowed by technical problem-solving. This role-aware orientation supports disciplined reflection and aligns development activity with research intent.

The framework further offers a principled approach to incorporating generative AI tools into research workflows. By positioning AI systems as collaborators that support, rather than replace, human judgement, SHAPR reinforces accountability and methodological responsibility. Explicit reflection on when and how AI assistance is employed helps ensure that the provenance of ideas, analyses, and artefacts remains transparent and defensible, addressing emerging concerns around authorship, attribution, and research integrity in AI-assisted development \citep{Amershi2019,Shneiderman2020}.

Beyond individual practice, SHAPR has implications for research supervision and coordination. The framework provides a shared conceptual language for discussing software development as part of research methodology, enabling more precise conversations about artefact evolution, reflective practice, and responsible AI use. In this way, SHAPR functions not only as guidance for individual researchers but also as a coordination mechanism that can reduce ambiguity and misalignment in supervisory relationships.

Overall, SHAPR encourages researchers to treat research software development as a legitimate site of knowledge production and learning. By embedding reflection, abstraction, and evidence management within development practice, the framework aligns technical work with scholarly inquiry and supports the generation of transferable design knowledge.

\subsection{Implications for learning and teaching}\label{discuss2}

In addition to its implications for research practice, SHAPR has important consequences for learning and teaching in software-intensive research contexts. By framing research software development as a structured, reflective, and human-centred activity, the framework aligns technical work with pedagogical goals related to deep learning, critical thinking, and methodological understanding \citep{Kennedy-Clark2013}. This alignment is particularly relevant in HDR education, where learners are expected to develop both technical competence and scholarly judgement.

A key pedagogical implication of SHAPR lies in its support for learning through structured making. Research software development often involves sustained engagement with complex and ill-defined problems, creating opportunities for experiential learning. However, without explicit scaffolding, learners may focus narrowly on implementation at the expense of reflection and conceptual understanding. SHAPR addresses this risk by embedding reflective checkpoints, role awareness, and artefact traceability into development practice, encouraging learners to connect technical activity with research aims and theoretical considerations \citep{Schon1983}.

The framework also supports metacognitive learning by making role transitions explicit. As solo researchers move between roles such as designer, developer, evaluator, and reflective analyst, SHAPR encourages awareness of how these shifts shape reasoning and decision-making. This role-aware perspective supports the development of transferable skills, including reflective reasoning, methodological planning, and responsible engagement with AI-assisted tools.

Generative AI introduces additional pedagogical challenges, which SHAPR addresses through its human-centred orientation. While AI-assisted tools can accelerate progress and lower technical barriers, they may also obscure learning processes if used uncritically. SHAPR positions AI systems as learning supports rather than learning substitutes, encouraging learners to interrogate AI-generated outputs and reflect on their appropriateness and limitations. This positioning aligns with broader educational concerns regarding meaningful learning in AI-augmented environments \citep{Shneiderman2020}.

From a teaching and supervision perspective, SHAPR offers a coherent framework for scaffolding discussions around research software development. Supervisors and educators can use the framework to guide feedback on artefact evolution, reflective practice, and methodological alignment, helping learners articulate not only what they have built, but how and why it contributes to their research. In this sense, SHAPR supports more consistent supervision and assessment practices in software-intensive research training.

\subsection{Limitations}\label{discuss3}

As a conceptual and design-oriented contribution, this study has several limitations that arise from its scope and purpose rather than from deficiencies in methodological rigour.

First, SHAPR is articulated and evaluated at a conceptual level. While the framework is grounded in established ADR principles and informed by literature on research software, solo development, and Human–AI collaboration, it is not empirically instantiated within a specific software development project in this paper. Consequently, claims regarding its effectiveness or impact remain analytical rather than empirical. This limitation is intentional and reflects the paper’s focus on establishing transferable design knowledge rather than reporting outcomes from a single instantiation \citep{GregorHevner2013}.

Second, the evaluation approach adopted is formative and reflective, emphasising internal coherence, methodological alignment, and contextual suitability. Although appropriate for framework development within ADR traditions \citep{Sein2011,Mullarkey2019}, this approach does not provide evidence of adoption or outcomes across diverse research settings. Empirical applications of SHAPR in future studies will be necessary to examine how the framework is enacted in practice and how it shapes development processes, learning trajectories, and research contributions.

Third, the framework is scoped specifically to solo research software development in educational and scholarly contexts. While some principles may be informative beyond these settings, SHAPR has not been designed or evaluated for large teams or industrial environments. Generalisation beyond the intended scope should therefore be undertaken cautiously.

Finally, although illustrative examples of AI-assisted tools are discussed, the framework reflects the state of AI-supported development practices at the time of writing. As tools and workflows evolve, specific examples may become outdated. However, because SHAPR is grounded in tool-agnostic, human-centred principles, its conceptual contribution is intended to remain relevant despite changes in the surrounding technological ecosystem.

Taken together, these limitations underscore the need for continued empirical and methodological work rather than detract from the contribution of the present study. By clearly delimiting its scope, this paper establishes SHAPR as a coherent foundation for cumulative research and learning in AI-assisted research software development.

\section{Conclusion}

This paper has addressed the growing need for structured support in solo, AI-assisted research software development, a context that is increasingly prevalent in Higher Degree Research (HDR) and educational settings yet remains under-theorised in existing methodological and pedagogical literature. While Action Design Research (ADR) provides a robust methodological foundation for studying and constructing socio-technical artefacts, it offers limited guidance on how its principles can be enacted in the day-to-day practice of solo researchers working with generative AI tools for software development.

To address this gap, the paper introduced the SHAPR framework—a Solo, Human-centred, AI-assisted PRactice—as a practice-level operational framework that complements ADR. SHAPR translates ADR’s methodological intent into actionable guidance by making explicit the roles, artefacts, reflective practices, and governance mechanisms required to sustain methodological rigour, human accountability, and learning in AI-assisted development contexts. By positioning SHAPR as design knowledge rather than as a new research methodology, the study preserves ADR’s epistemological foundations while extending its practical applicability to contemporary research environments.

The contribution of this paper is intentionally conceptual. The SHAPR framework itself constitutes the primary design artefact and unit of analysis and is evaluated through formative, reflective analysis of its internal coherence, alignment with ADR principles, and applicability to solo research practice. This positioning establishes a clear foundation for subsequent empirical application while avoiding premature claims regarding effectiveness or impact. In doing so, the paper demonstrates how design-oriented research can progress through staged contributions that move from framework articulation to applied investigation.

From the perspective of student learning, SHAPR foregrounds research software development as a site of learning, reflection, and knowledge construction rather than merely technical production. By embedding role awareness, artefact traceability, and human-centred use of generative AI within development practice, the framework supports deep learning and methodological understanding alongside technical competence. These qualities are particularly salient in research training contexts, where learners are developing both software artefacts and scholarly judgement.

Future work may apply SHAPR to the development of concrete research software artefacts across diverse domains, enabling empirical examination of how the framework shapes development processes, learning trajectories, and research contributions. Insights from such applications can inform further refinement of the framework through additional ADR cycles, contributing to a cumulative research programme on Human--AI collaboration and research software development. By articulating SHAPR as a coherent and transferable framework, this paper lays a conceptual foundation for that programme and contributes to ongoing discussions about how technology, knowledge, and learning intersect in AI-augmented research practice.

\end{document}